\documentclass[aps,prl,twocolumn,showpacs, amssymb, amsmath, aps, superscriptaddress] {revtex4-1}

\pdfoutput=1
\usepackage{ifpdf}

\usepackage{graphicx}% Include figure files
\usepackage{dcolumn}% Align table columns on decimal point
\usepackage{bm}% bold math
\usepackage{epstopdf}
\usepackage{latexsym}
\usepackage{mathrsfs}
\begin{document}

\title{Quantum control in the Cs $6S_{1/2}$ ground manifold using rf and ${\mu}$w magnetic fields}

\author{A. Smith}\altaffiliation[]{Current address: HRL Laboratories LLC,
3011 Malibu Canyon Road, Malibu, CA 90265.}\affiliation{Center for Quantum Information and Control, College of Optical Sciences and Department of Physics, University of Arizona, Tucson, AZ 85721, USA}
\author{B. E. Anderson}\affiliation{Center for Quantum Information and Control, College of Optical Sciences and Department of Physics, University of Arizona, Tucson, AZ 85721, USA}
\author{H. Sosa-Martinez}\affiliation{Center for Quantum Information and Control, College of Optical Sciences and Department of Physics, University of Arizona, Tucson, AZ 85721, USA}
\author{C. A. Riofr\'io}\altaffiliation[]{Current address: Dahlem Center for Complex Quantum Systems, Freie Universit\"at Berlin, 14195 Berlin, Germany.}\affiliation{Center for Quantum Information and Control, Department of Physics and Astronomy, University of New Mexico, Albuquerque, NM 87131, USA}
\author{Ivan H. Deutsch}\affiliation{Center for Quantum Information and Control, Department of Physics and Astronomy, University of New Mexico, Albuquerque, NM 87131, USA}
\author{Poul S. Jessen}\affiliation{Center for Quantum Information and Control, College of Optical Sciences and Department of Physics, University of Arizona, Tucson, AZ 85721, USA}

\date{\today}% It is always \today

\begin{abstract}
\noindent We implement arbitrary maps between pure states in the 16-dimensional Hilbert space associated with the ground electronic manifold of $^{133}$Cs. This is accomplished by driving atoms with phase modulated rf and $\mu$w fields, using modulation waveforms found via numerical optimization and designed to work robustly in the presence of imperfections. We evaluate the performance of a sample of randomly chosen state maps by randomized benchmarking, obtaining an average fidelity $>99\%$. Our protocol advances state-of-the-art quantum control and has immediate applications in quantum metrology and tomography.\end{abstract}


\maketitle

Coherent control of complex quantum systems plays an important and steadily increasing role across much of modern physics. In particular, high fidelity control is a cornerstone of quantum information science (QIS), where it is an essential part of quantum-enhanced approaches to computation \cite{Ladd2010}, simulation \cite{Bloch2012,Blatt2012,AspuruGuzik2012}, communication \cite{OBrien2009} and metrology \cite{Giovannetti2011}.  Because qubits are often encoded in physical spins, these tasks generally translate into control and measurement of individual and coupled spins. Atomic ground states, comprised of coupled nuclear and electronic spins, are a particularly attractive platform for QIS due to long coherence times and an existing, powerful toolbox for control and measurement. Examples include ion-trap quantum computers \cite{Wineland2011}, neutral-atom quantum simulators \cite{Bloch2012}, quantum memories \cite{Sangouard2011}, and spin squeezing  for quantum-limited clocks and magnetometers \cite{Ma2011}.
	
One of the most basic tasks of quantum control is to time-evolve a quantum system from a given initial to a desired final state (state mapping).  In this letter we explore the limits of state mapping between arbitrary pure states in a large Hilbert space, using as our test-bed the 16-dimensional hyperfine manifold associated with the electronic ground state of $^{133}$Cs atoms. The atomic evolution is driven by static, radio frequency (rf), and microwave ($\mu$w) magnetic fields, which is sufficient for full controllability in the entire ground manifold \cite{Merkel2008}. In contrast to past work based on the tensor light shift \cite{Chaudhury2007, Deutsch2010}, this approach is not affected by decoherence due to light scattering and associated optical pumping. As a result, our state map fidelities are limited only by imperfections in the applied magnetic fields, and we show that these can be compensated with ``robust" control techniques \cite{Mischuck2012} analogous to those used for spin-$1/2$ systems in nuclear magnetic resonance \cite{Vandersypen2005}. Finally, we implement and test a protocol for randomized benchmarking of state maps, inspired by those developed for Clifford gates in single- and few-qubit systems \cite{Knill2008,Ryan2009}. Combining these techniques, we have implemented and benchmarked a large sample of randomly chosen state maps and measured an average fidelity of $99.11(5)\%$. The corresponding infidelity is smaller by a factor of $5$ to $10$ relative to some recent experiments with similar-sized Hilbert spaces on other platforms \cite{Monz2011,Choi2010,Israel2012}, and thus represents a significant advance in state-of-the-art quantum control.  Such high-fidelity state mapping has important applications in quantum state preparation, e. g. known inputs for process tomography \cite{Chuang1997}, states that increase the coupling strength in atom-light interfaces and improve the generation of spin squeezing \cite{Norris2012}, and custom initial states for the study of non-equilibrium dynamics in spinor quantum gases \cite{Chang2005, Sadler2006, Liu2009}.

A detailed theoretical study of our scheme for quantum control of hyperfine-coupled electron and nuclear spins in alkali atoms can be found in \cite{Merkel2008}.  The most important conclusion of that work is that controllability can be achieved with a static bias magnetic field along $z$, combined with phase modulated rf magnetic fields along $x$ and $y$, and a phase modulated $\mu$w field driving a single transition between the hyperfine manifolds $F_\pm=I\pm1/2$. In this context, controllability means that the Hamiltonian dynamics can generate any transformation in $SU(d)$, where $d=2(2I+1)$ is the Hilbert space dimension of the alkali ground manifold for nuclear spin $I$.  In the case of $^{133}$ Cs we have $I=7/2$, and thus $F_{\pm}=3,4$ and $d=16$. In the rotating wave approximation, taking into account the finite nuclear magnetic moment and the second order Zeeman shift from the bias field, and driving the $\vert F=3,m=3\rangle \leftrightarrow \vert F=4,m=4\rangle$  $\mu$w transition, the corresponding control Hamiltonian has the form
\begin{equation*}
H_{\rm C} =H_0+H_{\rm rf}^{(3)}(\phi_x,\phi_y)+H_{\rm rf}^{(4)}(\phi_x,\phi_y)+H_{\mu {\rm w}}(\phi_{\mu {\rm w}}).
\label{eq:eq1}
\end{equation*}
For a derivation and the full form of this Hamiltonian, see the accompanying supplemental information and \cite{Smith2012}. We note that $H_0$ is independent of the control phases $\phi_x$, $\phi_y$ and $\phi_{\mu {\rm w}}$, that $H_{\rm rf}^{(3)}$ and $H_{\rm rf}^{(4)}$ are independent $SU(2)$ rotations of the $F_{\pm}=3,4$ manifolds controlled by the phases of the rf fields, and that $H_{\mu {\rm w}}$ is an $SU(2)$ rotation of the $\vert F_\pm,m=F_\pm\rangle$ pseudospin controlled by the phase of the $\mu$w field. Besides the control phases, the control Hamiltonian depends critically on an additional set of parameters $\Lambda=\{\Omega_0, \Omega_x, \Omega_y, \Omega_{\mu {\rm w}},\Delta_{\rm rf}, \Delta_{\mu {\rm w}}  \}$.  Here $\Omega_0=1$ MHz  is the Larmor frequency at which the spin ${\bf{F}}^{(4)}$ precesses in the bias field, $\Omega_x= \Omega_y=9$ kHz are the rf Larmor frequencies in the rotating frame, $\Omega_{\mu {\rm w}}=27.5$ kHz is the microwave Rabi frequency, and $\Delta_{\rm rf}= \Delta_{\mu {\rm w}}=0$ are the detunings of the rf and $\mu$w fields from resonance. As described below, our control fields are designed under the assumption that these parameters are very close to the indicated values; assuring that this is the case in the laboratory is one of the main challenges of the experiment. Details of how the parameters $\Lambda$ are measured and set to their design values, as well as how their spatial and temporal inhomogeneity are estimated, can be found in \cite{Smith2012}.

\begin{figure}
[t]\resizebox{8.25cm}{!}
{\includegraphics{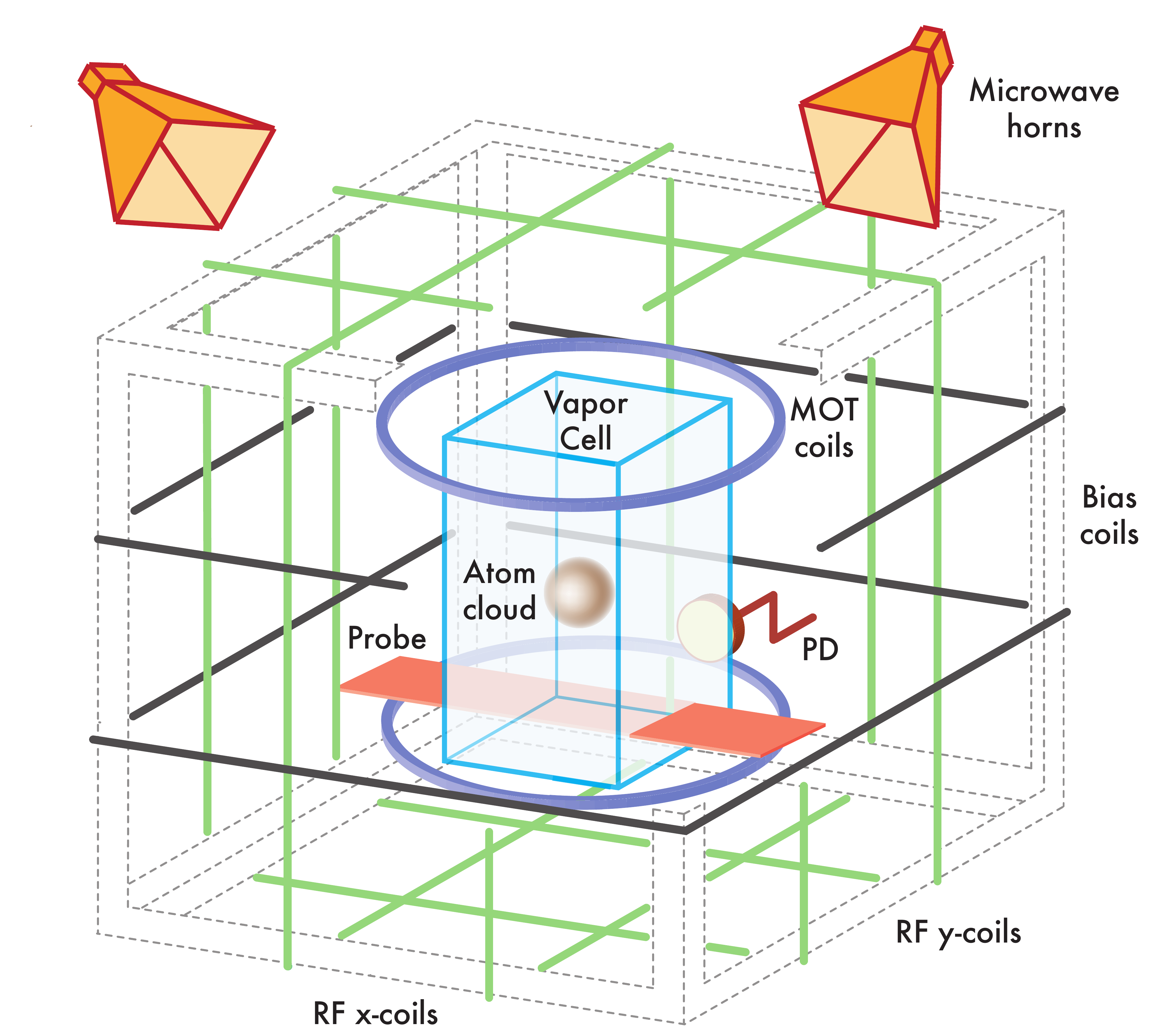}}
\caption{\label{fig:Fig1} (Color online) Schematic of the experimental setup.  Laser cooled Cs atoms are prepared in an all-glass vacuum cell centered within a plexiglas cube supporting the bias and rf coils. Microwave radiation is provided by two horn antennae.  Stern-Gerlach analysis is performed by letting the atoms fall in a magnetic field gradient provided by the MOT coils, and inferring the magnetic populations from the time dependent fluorescence excited by a probe beam and detected with a photodiode.}
\end{figure}

Our experimental setup (Fig.~\ref{fig:Fig1}) consists of a vapor-cell magneto-optic trap (MOT) and optical molasses, capable of preparing a few million Cs atoms at temperatures as low as $3$ $\mu$K.  The bias and rf magnetic fields are applied by three orthogonal coil pairs, each with a square cross section but otherwise close to Helmholtz configuration.  The DC current for the bias field is supplied by a modified, ultra-stable quasi-cw laser diode driver, while the current source for the rf fields is a dual-channel arbitrary waveform generator followed by power amplifiers. The microwave field is generated by a $\mu$w synthesizer running at $9.2$ GHz, mixed with a $30$ MHz signal from a arbitrary waveform generator, amplified and radiated by two separate microwave gain horns. The use of two gain horns results in significant improvement in the homogeneity of the $\mu$w intensity across the atom cloud. Using an all-glass vacuum cell and avoiding nearby conductive and magnetizable materials allows us to modulate the $1$ MHz rf fields in a bandwidth of a few hundred kHz.  Finally, synchronizing the experiment at a fixed point in the $60$ Hz AC powerline cycle allows us to measure and compensate static and AC background magnetic fields as described in \cite{Smith2011}.  As a result, our combined static bias and background fields along $z$ are accurate to $20$ ppm and stable to about $10$ ppm ($30$ $\mu$G).  The bias field along $z$ makes the presence of background fields along $x$ and $y$ less critical, and only static compensation at the mG level is required here.

An experimental sequence begins by releasing a cold atom sample into free fall. We use a combination of optical pumping and Larmor precession to initialize the atoms in a fiducial state $\vert F=4,m=4\rangle$, at which point the static bias field is switched on to maintain the orientation of the spin.  The bias field stabilizes to the required $10$ ppm level in $\sim 7$ ms, at which point we apply rf and $\mu$w fields with predetermined phase modulation waveforms over a time $T$ to evolve the spins until they closely approach the desired target state. Finally we measure the populations in the $16$ magnetic sublevels $\vert F,m\rangle$, by performing Stern-Gerlach analysis as described in \cite{Klose2001} and detecting atoms in the $F_{\pm}$ manifolds with separate optical probe beams.  

\begin{figure}
[t]\resizebox{8.5cm}{!}
{\includegraphics{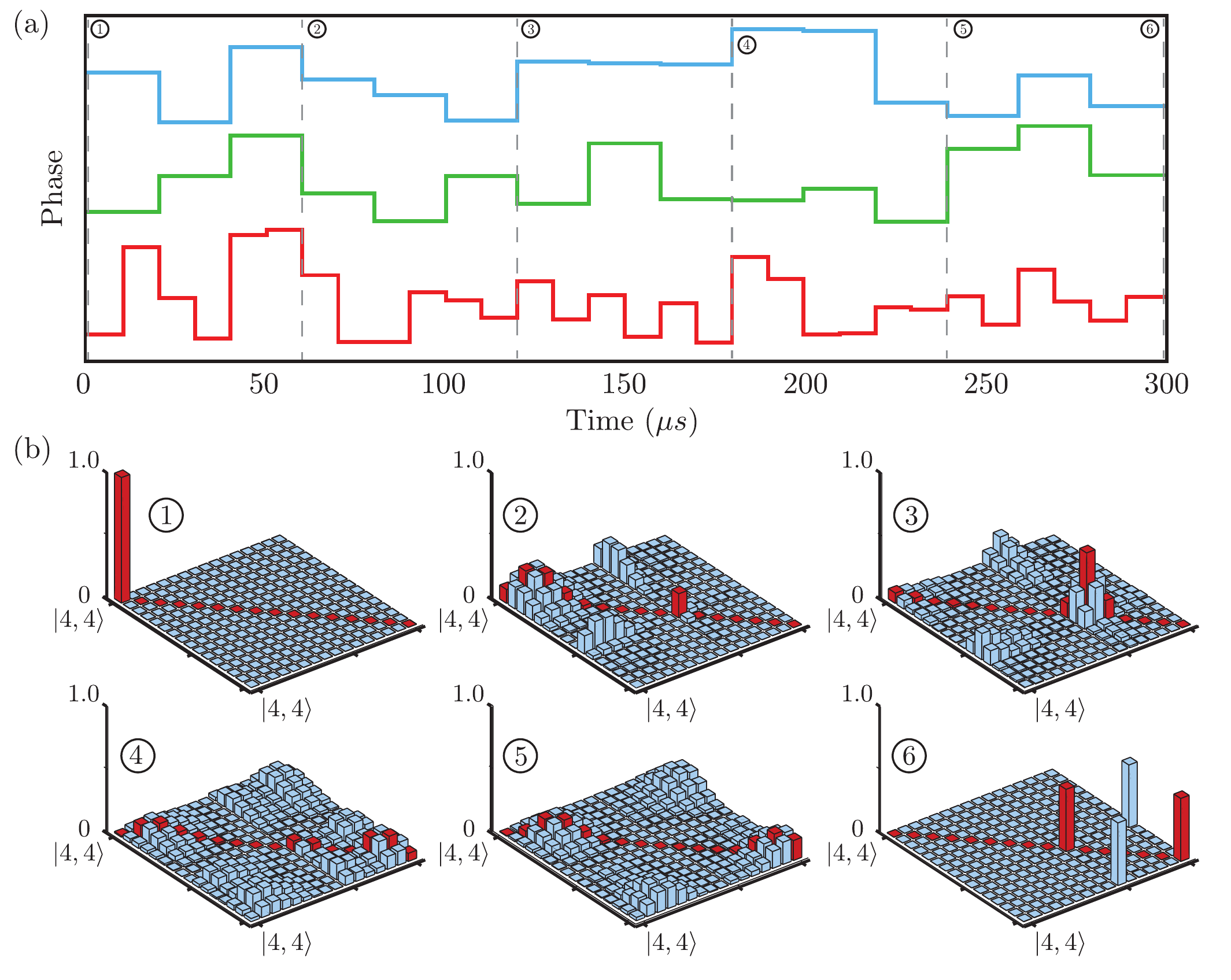}}
\caption{\label{fig:Fig2} (Color online) Implementation of the quantum state map $\vert4,4\rangle\rightarrow(\vert3,3\rangle+\vert3,-3\rangle)/\sqrt{2}$. (a) Phase modulation waveform for the rf (top, middle) and $\mu$w (bottom) fields. (b) Numerical simulation of the evolving quantum state, shown as density matrices for the times indicated. Populations are shown as dark (red) and coherences (absolute values only) as light (blue) tones. Magnetic sublevels are ordered $\{\vert4,4\rangle,...,\vert4,-4\rangle,\vert3,3\rangle,...,\vert3,-3\rangle\}$ along the axes.}
\end{figure}

Control fields that accomplish a given state map are found using numerical techniques common to optimal control.  Starting from some initial state, the goal is to find a set of time dependent phases $\{\phi_x(t),\phi_y(t),\phi_{\mu {\rm w}}(t) \}$ such that the fidelity relative to the target state, $\mathcal{F}=\vert \langle \psi_{\rm target} \vert \psi(T) \rangle \vert^2$, is maximized after a fixed control time $T$. Maximization is done with a gradient ascent algorithm,  where in each iteration the time-evolved state $\vert \psi(T)\rangle$ is found by numerical integration of the Sch\"odinger equation, starting from $\vert \psi_{\rm initial} \rangle$ and using the given values of the phases.  To increase the speed of integration we keep the phases piece-wise constant in time, typically using $30$ time steps for the $\mu$w phase and $15$ time steps for each rf phase in a ``control waveform" of $300$ $\mu$s duration (Fig.~\ref{fig:Fig2}a).  The total number of control variables ($60$) is thus well above the $2d-2=30$ real-valued parameters required to specify the transformation $\vert\psi_{\rm initial}\rangle\rightarrow\vert\psi_{\rm target} \rangle$. We begin the numerical search for phases $\{\phi_x^{(i)},\phi_y^{(i)},\phi_{\mu {\rm w}}^{(i)} \}$ with a random guess, and then use a standard routine from the MatLab optimization toolbox to iteratively maximize $\mathcal{F}$. The result is a control waveform corresponding to a local maximum in the control landscape; it is our experience that different initial guesses lead to different control waveforms, but that if $T$ is large enough nearly every initial guess will result in a waveform that achieves $>99\%$ fidelity. This is consistent with the expected benign nature of the search landscape \cite{Rabitz2004}.

The optimization procedure can be extended to find control waveforms that are robust in the presence of errors and imperfections.  In our case the dominant imperfections are spatial inhomogeneities and shot-to-shot variations of the parameters in $H_{\rm C}$.  A robust control waveform can then be found by maximizing the average fidelity $\bar{\mathcal{F}}=\int_\Lambda P(\Lambda) \vert \langle \psi_{\rm target} \vert \psi_\Lambda(T) \rangle \vert^2 d\Lambda $, where $P(\Lambda)$ is the probability that the parameters take on values $\Lambda$, and $\vert \psi_\Lambda(T) \rangle$ is the corresponding final state \cite{Kobzar2005}.  In practice we have found it sufficient to average over three values of the bias field, $\{\Omega_0,\Omega_0\pm\delta\Omega_0\}$, and three values of the $\mu$w Rabi frequency, $\{\Omega_{\mu {\rm w}},\Omega_{\mu {\rm w}}\pm\delta\Omega_{\mu {\rm w}}\}$, for a total of nine combinations of parameter values.  For simplicity we assume each combination is equally probable, and use variations $\delta\Omega_0=100$ Hz and $\delta\Omega_{\mu {\rm w}}=140$ Hz that are slightly larger than our estimated standard deviations. This relatively coarse sampling of the probability distribution speeds optimization, and we have found that the resulting, optimized control waveform performs well when its fidelity is averaged using a finer sampling of the estimated Gaussian distributions. Again, it is our experience that waveforms with fidelity in excess of $99\%$ can almost always be found from a single initial guess.  Figure ~\ref{fig:Fig2}b illustrates the performance of a robust control waveform designed in this fashion. The figure shows intermediate and final density matrices from a numerical simulation that includes an average over $\Lambda$, with conservative estimates for the uncertainty of every parameter. The resulting final state is very slightly mixed, but the state map fidelity remains very high.

\begin{figure}
[t]\resizebox{8.25cm}{!}
{\includegraphics{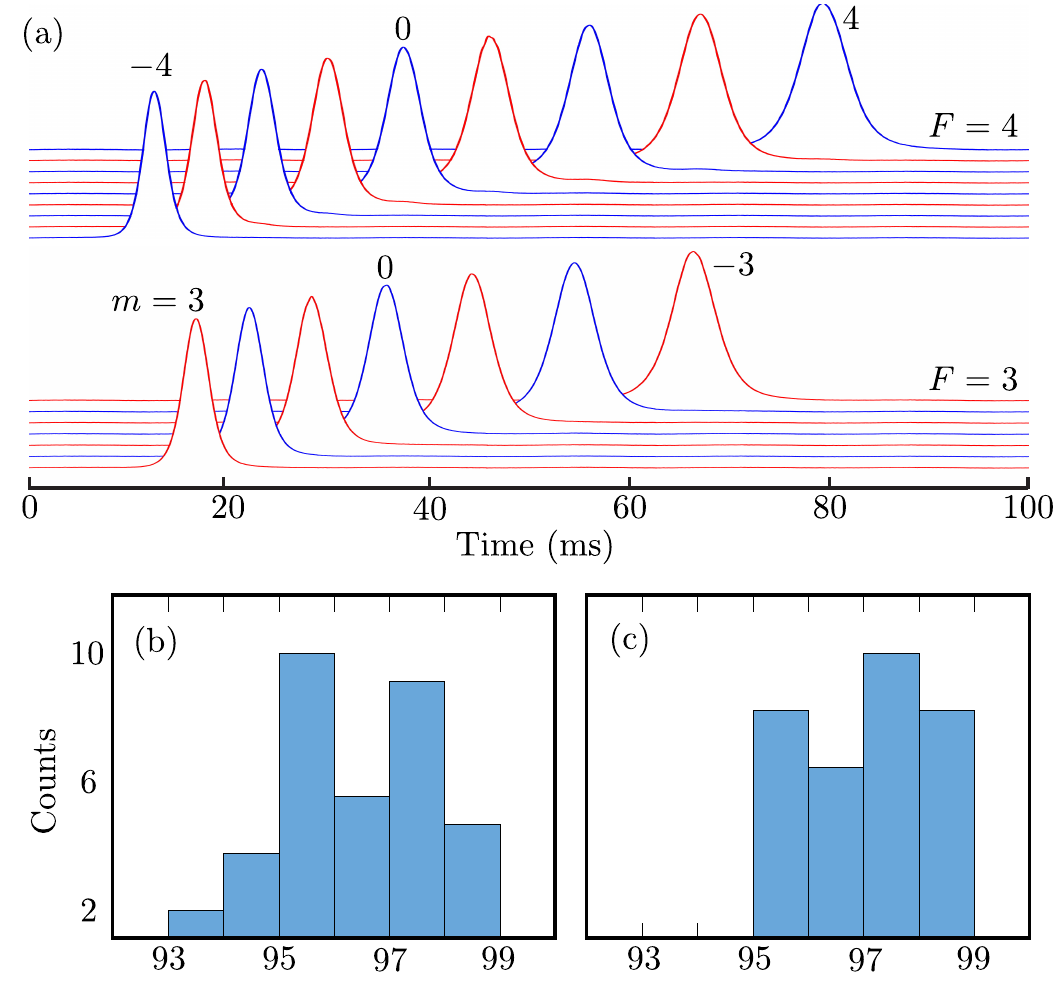}}
\caption{\label{fig:Fig3} (Color online) (a) Stern Gerlach analysis of magnetic sublevel populations, in the form of arrival time distributions at the probe beam. Each line is a separate measurement after a state map $\vert 4,4 \rangle  \rightarrow \vert F,m \rangle$ as indicated. (b) Histogram of the fidelities for 32 non-obust state maps of this form. (c) Histogram of the fidelities for 32 robust state maps of this form.}
\end{figure}

The simplest experimental test of our state mapping protocol is to start from $\vert F=4,m=4 \rangle$, map to any one of the states $\vert F,m \rangle$, and estimate the fidelity directly by measuring the population of the target state by Stern-Gerlach analysis.  Figure ~\ref{fig:Fig3}a shows Stern-Gerlach signals for maps to each of the $16$ magnetic sublevels in the ground manifold, while Figs. ~\ref{fig:Fig3}b\&c show histograms of the estimated fidelity for $32$ non-robust and $32$ robust control waveforms (the sets contain two different control waveforms for each map $\vert 4,4 \rangle \rightarrow \vert F,m \rangle$).  The trend in this data suggests that robust waveforms slightly outperforms non-robust waveforms. However, the estimated fidelities include a substantial contribution from errors in initial state preparation and final state readout, and are therefore not an accurate measure of the fidelity of the state maps themselves.  Furthermore, this simple technique cannot be used to estimate the fidelity of state maps where the final state is a coherent superposition of two or more magnetic sublevels. In \cite{Smith2013} we used the state mapping procedure discussed here to produce complex input states for tomography, and comparisons between a few (relatively low fidelity) reconstructions and the corresponding target states can be seen there.

\begin{figure}
[t]\resizebox{8.25cm}{!}
{\includegraphics{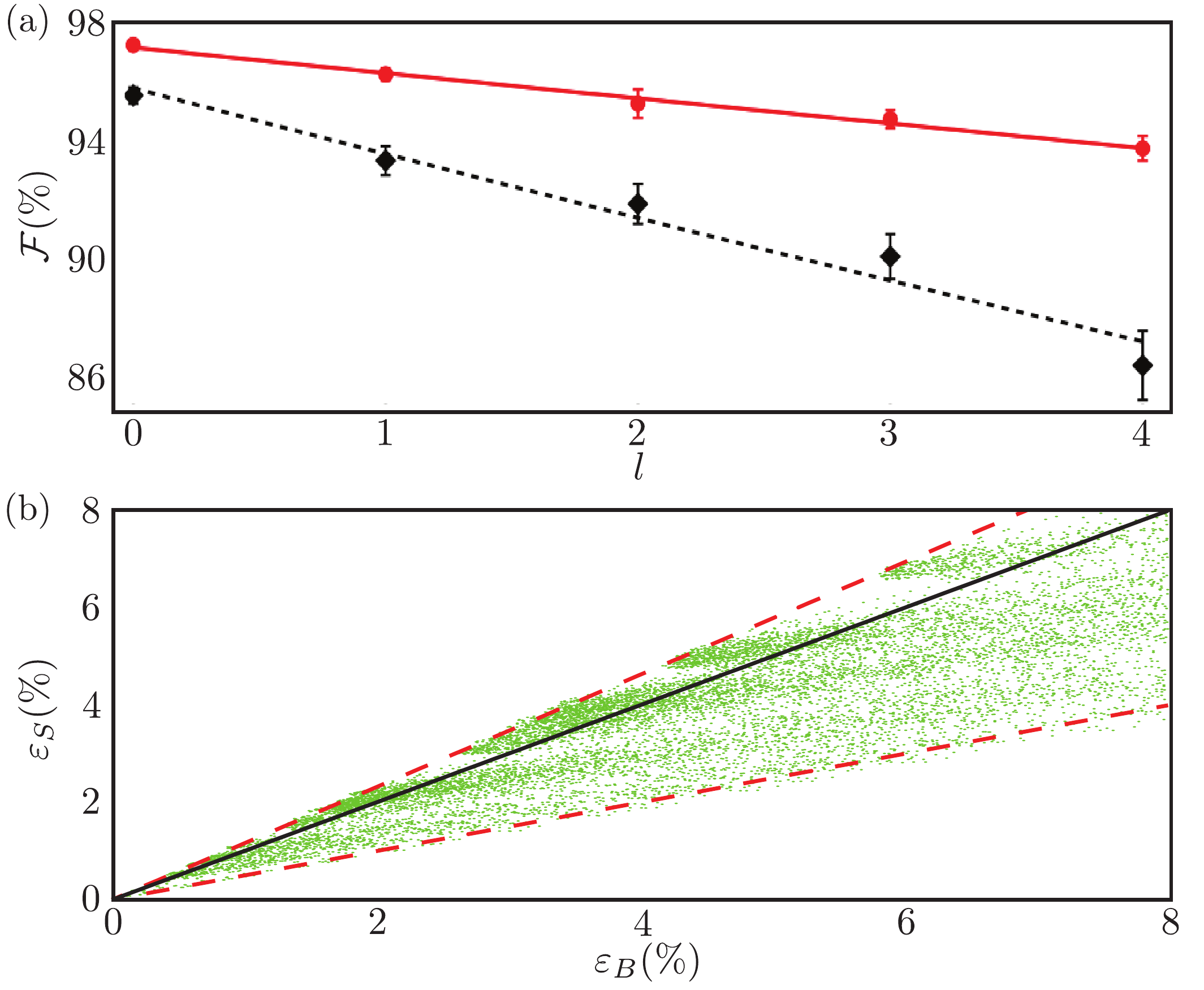}}
\caption{\label{fig:Fig4} (Color online) (a) Randomized benchmarking data showing the overall fidelity for sequences of up to $4$ state maps.  Points are experimental data, lines are fits of the form $\mathcal{F}(l)$, for robust (red circles) and non-robust (black diamonds) control waveforms. Error bars are one standard deviation for the average over different state map sequences. The average fidelity per state map inferred from the fits are $99.11(5)\%$ and $97.7(3)\%$ respectively. (b) Plot showing the correlation between benchmarking and standard fidelities.  Each data point $\left( \epsilon_B,\epsilon_S \right)$ is obtained from a numerical simulation performed with a distinct set of values for the parameters in $H_C$. Solid and dashed lines correspond to $\epsilon_S=\epsilon_B$, $\epsilon_S=0.5 \epsilon_B$, and $\epsilon_S=1.15 \epsilon_B$, respectively.}
\end{figure}

To obtain an accurate measure of state map fidelity we employ a procedure inspired by the randomized benchmarking protocol developed for single- and multi-qubit Clifford gates \cite{Knill2008,Ryan2009}.  The basic idea is to apply state maps in progressively longer sequences, i. e.
\begin{align*}
&\vert 4,4 \rangle \rightarrow \vert \psi_0 \rangle \rightarrow \vert 4,4 \rangle, \\
&\vert 4,4 \rangle \rightarrow \vert \psi_0 \rangle \rightarrow \vert \psi_1 \rangle \rightarrow \vert 4,4 \rangle, \\
&\vert 4,4 \rangle \rightarrow \vert \psi_0 \rangle \rightarrow \ldots  \rightarrow \vert \psi_l \rangle \rightarrow \vert 4,4 \rangle,
\end{align*}
and estimate the overall fidelity of each sequence by measuring the population returned to $\vert 4,4 \rangle$.  To increase sample size we consider a number of such progressions, each consisting of different sequences with intermediate states $\vert \psi_0 \rangle,\ldots,\vert \psi_l \rangle$ chosen at random according to the Haar measure \cite{Mezzadri2007}.  For each progression we design control waveforms to perform the corresponding state maps, implement these in the laboratory, and measure the overall fidelity as function of $l$.  Finally we average together the fidelities observed for the different progressions, which improves statistics and smooths out fluctuations from accidental spin-echo effects in the individual progressions.  The resulting data is fit to a function
\begin{equation*}
\mathcal{F}(l)=\frac{1}{d}+\frac{d-1}{d}\left(1-\frac{d}{d-1}\epsilon_0\right)\left(1-\frac{d}{d-1}\epsilon\right)^l,
\label{eq:controlH}
\end{equation*}
where $d=16$ is the Hilbert space dimension, $\epsilon$ is the average error per state map, and $\epsilon_0$ is the average combined error in the preparation (optical pumping into $\vert 4,4 \rangle$ and mapping $\vert 4,4 \rangle \rightarrow \vert \psi_0 \rangle$) and readout (mapping $\vert \psi_l \rangle \rightarrow \vert 4,4 \rangle$ and measuring the $\vert 4,4 \rangle$ population) steps.  This generalization of the fit function used for qubits \cite{Ryan2009} ensures proper asymptotic behavior for large and small $l$.  Figure ~\ref{fig:Fig4}a shows typical data from this randomized benchmarking protocol, from which we infer a fidelity per state map $\mathcal{F}=1-\epsilon$ of $99.11(5)\%$ and $97.7(3)\%$ for robust and non-robust control waveforms, respectively.

As a final step, we use numerical modeling to check that our benchmarking protocol yields average fidelities in reasonable agreement with other measures.  We do this in two steps, first by generating simulated benchmarking data analogous to Fig.~\ref{fig:Fig4}a and fitting it to obtain average state map errors $\epsilon_B$, and secondly, by calculating the standard infidelities $1-\vert \langle \psi_{\rm target} \vert \psi(T) \rangle \vert^2$ for the state maps used in the simulation and averaging those to obtain an  average standard error $\epsilon_S$. This process is repeated for many parameter values $\Lambda$, each time producing a data point $\left( \epsilon_B,\epsilon_S \right)$ for the possible correlation between the two measures. Figure ~\ref{fig:Fig4}b shows a large collection of such data points for parameters $\Lambda$ that go well beyond the range likely to be present in our experiment. If our benchmarking protocol is reasonable one would expect all those data points to lie near the line $\epsilon_S=\epsilon_B$. In practice they appear to fall mostly below that line, clustered roughly in the range $0.5 \epsilon_B < \epsilon_S <1.15\epsilon_B$  This suggests that in some situations the benchmarking protocol may overestimate the standard error by as much as a factor of two.  In the context of our experiment, this means the average of the standard fidelity for a set of randomly chosen state maps is likely to lie between $99\%$ and $99.5\%$, a result that could not easily have been established by other means.

In conclusion, we have demonstrated that high fidelity quantum state mapping can be implemented in the $16$-dimensional hyperfine ground manifold of $^{133}$Cs, by driving the system solely with phase modulated rf and $\mu$w magnetic fields.  Robust controls can be efficiently designed to compensate for imperfections in the driving fields, leading to significant improvements in the accuracy of the state maps. A randomized benchmarking protocol was implemented and showed that the average fidelity of such robust state maps is $99\%$ or greater.  Future use of this platform includes the exploration of control tasks that are more complex than state maps, e. g., unitary transformations on the entire ground manifold or subspaces thereof, and partial isometries that map between subspaces.  Such studies will help address questions related to the feasibility of numerical search for control waveforms that implement those types of transformations \cite{Moore2011}, as well as the possibility of inhomogeneous control and whether control can be robust in the presence of static and time dependent perturbations \cite{Mischuck2012}.

This work was supported by the US National Science Foundation Grants PHY-1212308, -1212445, -0903930, and -0969997.

% Comment out next lineafter creating the compiled bbl file and then cut&paste from bbl file

\newpage
\onecolumngrid

% avoids incorrect hyphenation, added Nov/08 by SSR
\hyphenation{ALPGEN}
\hyphenation{EVTGEN}
\hyphenation{PYTHIA}

\section{SUPPLEMENTARY MATERIAL}

\section{Cs atoms in static, radio frequency and microwave magnetic fields}
\noindent The hyperfine Hamiltonian for an alkali atom in the presence of a magnetic field is
\begin{equation}
\label{eq:BohrMagneticHammy}
H = A \bf{I} \cdot \bf{S} + \it{g}_{\rm S} {\mu}_{\rm B} \bf{S} \cdot \bf{B} + \it{g}_{\rm I} {\mu}_{\rm B} \bf{I} \cdot \bf{B},
\end{equation}
where $g_{\rm S}$ and $g_{\rm I}$ are the electron and nuclear $g$-factors respectively.  When the magnetic interaction is negligible compared to the hyperfine interaction, $\mu_{\rm B} |\bf{B}| \ll \it{A}$, Eq. \ref{eq:BohrMagneticHammy} can be rewritten in terms of operators that act separately in the $F_{\pm}=3,4$ manifolds,
\begin{equation}
\label{eq:SubspacesMagneticHammy}
H = \frac{\Delta E_{\rm HF}}{2}(P^{(4)}-P^{(3)}) +{ \it{g}}_4{\mu}_{\rm B} {\bf{F}}^{(4)} \cdot {\bf{B}} + {\it{g}}_3{\mu}_{\rm B} {\bf{F}}^{(3)} \cdot \bf{B}.
\end{equation}\\
Here $\Delta E_{\rm HF}$ is the hyperfine splitting, and  $P^{(\pm)}$, ${\bf{F}}^{(\pm)} = P^{(\pm)} {\bf{F}} P^{(\pm)}$, and $g_{\pm}$ are the projectors, angular momenta, and Land\'e $g$-factors associated with the $F_{\pm}$ manifolds, respectively.\\

In our experiment the magnetic field  ${\bf B}(t)=B_0{\bf e}_z+{\bf B}_{\rm{rf}}(t)+{\bf B}_{\mu \rm{w}}(t)$, where the magnitude of the static bias field far exceeds that of the rf and $\mu$w components. Even though our system is deep in the linear Zeeman regime, ${\mu}_{\rm B} B_0<<A$, a second order correction to Eq. \ref{eq:SubspacesMagneticHammy} is necessary to model the dynamics with sufficient accuracy.  This can be done using the Breit-Rabi solution for the energies of the magnetic sublevels in the presence of the bias field only,

\begin{equation}
\label{eq:BreitRabiFormula}
E(m_{\pm}) = -\frac{\Delta E_{\rm HF}}{16} + m_{\pm} \bar{x} \pm \frac{\Delta E_{\rm HF}}{2} \sqrt{1 + \frac{4 m_{\pm}}{8} x + x^{2}},
\end{equation}  \\
where $m_{\pm}$ are the magnetic quantum numbers in the $F_{\pm}$ manifolds, and $\bar{x} = g_{\rm I} \mu_{\rm B} B$ and  $x = \frac{\mu_{\rm B} B}{\Delta E_{\rm HF}}(g_{\rm S} - g_{\rm I})$ are the Breit-Rabi variables.  Keeping terms up to second order in $x$ and substituting in Eq. \ref{eq:SubspacesMagneticHammy}, we get the static part of the Hamiltonian including the bias field,

\begin{eqnarray}
\label{eq:BreitRabiFormulaApprox2}
H_{0} &=& \bigg (\frac{\Delta E_{\rm HF}}{2} + \frac{\Delta E_{\rm HF} x^2}{4}\bigg ) \big (P^{(4)}-P^{(3)} \big) + \bigg (\frac{\Delta E_{\rm HF} x}{8} + \bar{x} \bigg ) F_{z}^{(4)} \nonumber \\
& & -\bigg (\frac{\Delta E_{\rm HF} x}{8} - \bar{x}\bigg ) F_{z}^{(3)} - \frac{\Delta E_{\rm HF} x^2}{64} \big ({F_{z}^{(4)}}^{2} - {F_{z}^{(3)}}^{2}\big ).
\end{eqnarray}\\
Substituting $ \hbar\Omega_{0} = (\frac{\Delta E_{\rm HF} x}{8} + \bar{x})$ and $g_{\rm rel} = g_3/g_4$, this can be written

\begin{eqnarray}
\label{eq:QuadraticZeemanHammy}
H_{0} &=& \bigg(\frac{\Delta E_{\rm HF}}{2} - \frac{16 g_{\rm rel} \hbar^{2} \bar{\Omega}_{0}^{2}}{\Delta E_{\rm HF}}\bigg )\big (P^{(4)}-P^{(3)}\big ) + \hbar\Omega_{0}\big (F_{z}^{(4)} + g_{\rm rel} F_{z}^{(3)}\big ) \nonumber \\
& & +\frac{g_{\rm rel} \hbar^{2} \bar{\Omega}_{0}^{2}}{\Delta E_{\rm HF}}\big ({F_{z}^{(4)}}^{2} - {F_{z}^{(3)}}^{2}\big ),
\end{eqnarray}\\
where  $\hbar^2 \bar{\Omega}_{0}^2=\hbar^2 {\Omega}_{0}^2-4\bar{x}^2/g_{\rm{rel}}$.  For our purpose the approximation $\bar{\Omega}_{0}\approx {\Omega}_{0}$ is sufficient throughout.  Note that the difference in $g$-factors for the $F_{\pm}$ hyperfine manifolds, $g_{\rm rel} = -1.0032$, cause the two manifolds to precess in opposite directions and at slightly different rates.  Furthermore, the quadratic Zeeman term, while small, can lead to qualitative changes in the spin dynamics, e. g. collapse and revival of the mean spin.\\

We next consider the addition of rf and $\mu$w fields. Our rf magnetic field has two orthogonal components, ${\bf B}_{\rm{rf}}(t)=B_x {\rm{cos}}(\omega_{\rm{rf}}t-\phi_x){\bf e}_x+B_y {\rm{cos}}(\omega_{\rm{rf}}t-\phi_y){\bf e}_y$, and the Hamiltonian for the rf interaction is

\begin{equation}
\label{rfHammy}
H_{\rm rf} =\hbar\Omega_{x}{\rm cos}(\omega_{\rm rf} t -\phi_{x})(F_{x}^{(4)} + g_{\rm rel} F_{x}^{(3)})+\hbar \Omega_{y}{\rm cos}(\omega_{\rm rf} t -\phi_{y})(F_{y}^{(4)} + g_{\rm rel} F_{y}^{(3)}),
\end{equation}\\
where $\hbar \Omega_{x} = g_4 \mu_{\rm B} B_{x}$ and $\hbar \Omega_{y} = g_4 \mu_{\rm B} B_{y}$. Our $\mu$w magnetic field, ${\bf B}_{\mu{\rm w}}(t)={\bf B}_{\mu{\rm w}}{\rm cos}(\omega_{\mu{\rm w}}t-\phi_{\mu{\rm w}})$ is resonant with the $\left | F=4,m=4 \right \rangle \rightarrow \left | F=3,m=3 \right \rangle$ ``stretched state" transition, and the bias field ensures that all other transitions are off-resonance by at least $\Omega_{0}$ and can be ignored.  The Hamiltonian for the microwave interaction can then be written as
\begin{equation}
\label{microwaveHammy}
H_{\mu {\rm w}} = \hbar \Omega_{\mu {\rm w}}{\rm cos}(\omega_{\mu {\rm w}} t -\phi_{\mu {\rm w}}) \big( \! \left | 4,4 \right \rangle \langle 3,3  | + \left | 3,3 \right \rangle \langle 4,4  | \big),
\end{equation}\\
where the Rabi frequency $\Omega_{\mu {\rm w}}$ depends on the intensity and polarization of the $\mu$w field.

The overall  Hamiltonian, $H = H_{0}+ H_{\rm rf}+H_{\mu {\rm w}}$, can be recast in a more useful form, by a suitable transformation followed by a rotating wave approximation (RWA).  The transformed Hamiltonian is of the form

\begin{equation}
\label{hammyTransform}
H' = U^{\dagger}HU -i \hbar U^{\dagger}\frac{dU}{dt} 
= U^{\dagger}H_{0}U - i \hbar U^{\dagger}\frac{dU}{dt}+ U^{\dagger}H_{\rm rf}U+U^{\dagger}H_{\mu {\rm w}}U .
\end{equation}
\noindent where
\begin{equation}
\label{frameTransformation}
U = e^{\frac{-i \alpha t}{2}(P^{(4)}-P^{(3)})} e^{-i \omega_{\rm rf} t (F^{(4)}_{z}-F^{(3)}_{z})},  \hspace{10mm} \alpha=\omega_{\mu\rm{w}}-7\omega_{\rm{rf}}.
\end{equation}\\
\noindent Considering first the static part of the Hamiltonian, $H_{0}' =U^{\dagger}H_{0}U -i \hbar U^{\dagger}dU/dt$, and noting that the operators $P^{(\pm)}$, $F_z^{(\pm)}$, $U$ and $H_0$ all commute, one easily finds

\begin{equation}
H_{0}' = H_0-\frac{\hbar\alpha}{2} \big(P^{(4)}-P^{(3)}\big)-\hbar \omega_{\rm rf} \big(F^{(4)}_{z}-F^{(3)}_{z}\big).
\end{equation}\\
\noindent It is useful to rewrite this in terms of rf and $\mu$w detunings, $\Delta_{\rm{rf}}=\omega_{\rm{rf}}-\Omega_0$ and $\Delta_{\mu {\rm w}} =\omega_{\mu {\rm w}} - \omega_{\left | 4,4 \right \rangle \rightarrow \left | 3,3 \right \rangle}$, where $\omega_{\left | 4,4 \right \rangle \rightarrow \left | 3,3 \right \rangle} = \Delta E_{\rm HF}/\hbar - 7 g_{\rm rel} \hbar \Omega_{0}^{2} / \Delta E_{\rm HF} + (4-3g_{\rm rel})\Omega_{0}$ is the splitting between the $\left |4,4\right \rangle$ and $\left |3,3\right \rangle$ states in the presence of the static bias field.  This gives us

\begin{align}
\label{staticHammyTransform}
\nonumber
H_{0}' =& \;  \Big\lbrack \frac{3 \hbar \Omega_{0}}{2}(1+g_{\rm rel}) - \frac{25 g_{\rm rel} \hbar^{2} \Omega_{0}^{2}}{2 \Delta E_{\rm HF}} - \frac{\hbar}{2}(\Delta_{\mu {\rm W}} - 7 \Delta_{\rm rf}) \Big\rbrack \big(P^{(4)}-P^{(3)}\big) \nonumber \\ 
& + \hbar\Omega_{0}(1+g_{\rm rel})F_{z}^{(3)} + \frac{g_{\rm rel} \hbar^2\Omega_{0}^{2}}{\Delta E_{\rm HF}}\big({F_{z}^{(4)}}^{2} - {F_{z}^{(3)}}^{2}\big) - \hbar\Delta_{\rm rf} \big(F_{z}^{(4)} - F_{z}^{(3)}\big).
\end{align}\\
Transformation of the rf part of the Hamiltonian, $H_{\rm rf}' =U^{\dagger}H_{\rm rf}U$, can be accomplished by noting that 

\begin{equation}
e^{\frac{i \alpha t}{2}(P^{(4)}-P^{(3)})}F^{(\pm)}_{x,y}e^{\frac{-i \alpha t}{2}(P^{(4)}-P^{(3)})}  = F^{(\pm)}_{x,y},
\end{equation}
and
\begin{align}
&e^{i \omega_{\rm rf} t (F^{(4)}_{z}-F^{(3)}_{z})} F^{(\pm)}_{x} e^{-i \omega_{\rm rf} t (F^{(4)}_{z}-F^{(3)}_{z})} =F^{(\pm)}_{x} {\rm cos}(\omega_{\rm rf} t)\mp F^{(\pm)}_{y} {\rm sin}(\omega_{\rm rf} t), \nonumber \\
e&^{i \omega_{\rm rf} t (F^{(4)}_{z}-F^{(3)}_{z})} F^{(\pm)}_{y} e^{-i \omega_{\rm rf} t (F^{(4)}_{z}-F^{(3)}_{z})} =\pm F^{(\pm)}_{x} {\rm sin}(\omega_{\rm rf} t)+ F^{(\pm)}_{y} {\rm cos}(\omega_{\rm rf} t),
\end{align}
\noindent which gives us

\begin{align}
\label{rfHammyTransform}
H_{\rm rf}'=&  \; \hbar\Omega_{x}{\rm cos}(\omega_{\rm rf}t - \phi_{x}) \Big ( F^{(4)}_{x} {\rm cos}(\omega_{\rm rf}t) - F^{(4)}_{y} {\rm sin}(\omega_{\rm rf}t)+ g_{\rm rel}F^{(3)}_{x} {\rm cos}(\omega_{\rm rf}t) + g_{\rm rel}F^{(3)}_{y} {\rm sin}(\omega_{\rm rf}t)\Big ) \nonumber \\
& +  \hbar\Omega_{y}{\rm cos}(\omega_{\rm rf}t - \phi_{y}) \Big ( F^{(4)}_{x} {\rm sin}(\omega_{\rm rf}t) + F^{(4)}_{y} {\rm cos}(\omega_{\rm rf}t) - g_{\rm rel}F^{(3)}_{x} {\rm sin}(\omega_{\rm rf}t) + g_{\rm rel}F^{(3)}_{y} {\rm cos}(\omega_{\rm rf}t)\Big ).
\end{align}\\
\noindent Multiplying through and dropping terms that oscillate at frequency $2\omega_{\rm rf}$, we obtain the rf Hamiltonian in the rotating wave approximation,

\begin{align}
H_{\rm rf}'=& \;  \frac{\hbar\Omega_{x}}{2} \Big\lbrack {\rm cos}(\phi_{x})F^{(4)}_{x} - {\rm sin}(\phi_{x})F^{(4)}_{y}\Big\rbrack + \frac{\hbar\Omega_{y}}{2}\Big\lbrack {\rm sin}(\phi_{y})F^{(4)}_{x}+{\rm cos}(\phi_{y})F^{(4)}_{y} \Big\rbrack \nonumber \\
& + \frac{g_{\rm rel}\hbar\Omega_{x}}{2} \Big\lbrack {\rm cos}(\phi_{x})F^{(3)}_{x} + {\rm sin}(\phi_{x})F^{(3)}_{y}\Big\rbrack + \frac{g_{\rm rel}\hbar\Omega_{y}}{2}\Big\lbrack -{\rm sin}(\phi_{y})F^{(3)}_{x}+{\rm cos}(\phi_{y})F^{(3)}_{y} \Big\rbrack \nonumber \\
& = H_{\rm{rf}}^{(4)}(\phi_x,\phi_y)+H_{\rm{rf}}^{(3)}(\phi_x,\phi_y).
\end{align}\\
\noindent Finally, the $\mu$w part of the Hamiltonian transforms as

\begin{align}
\label{microwaveHammyTransform}
H_{\mu {\rm w}}' =& \; U^{\dagger}H_{\mu {\rm w}}U \nonumber \\
=& \; \hbar \Omega_{\mu {\rm w}}{\rm cos}(\omega_{\mu {\rm w}} t -\phi_{\mu {\rm w}})e^{\frac{i \alpha t}{2}(P^{(4)}-P^{(3)})}e^{i \omega_{\rm rf} t (F^{(4)}_{z}-F^{(3)}_{z})}   \nonumber \\
 &  \times \big( \! \left | 4,4 \right \rangle \langle 3,3  | + \left | 3,3 \right \rangle \langle 4,4  | \big) e^{\frac{-i \alpha t}{2}(P^{(4)}-P^{(3)})} e^{-i \omega_{\rm rf} t (F^{(4)}_{z}-F^{(3)}_{z})}. 
 \end{align}
\noindent Using 
\begin{equation}
e^{i\frac{\alpha t}{2}(P^{(4)}-P^{(3)})}\vert F^{(\pm)},m\rangle=e^{\pm i\frac{\alpha t}{2}}\vert F^{(\pm)},m\rangle, \nonumber
\end{equation}
\vspace{-7mm}
\begin{equation}
e^{i\omega_{\rm{rf}}(F_z^{(4)}-F_z^{(3)})}\vert F^{(\pm)},m\rangle=e^{\pm i m\omega_{\rm{rf}}}\vert F^{(\pm)}m\rangle,
\end{equation}

\noindent we can rewrite this as 

\begin{align}
\label{microwaveHammyTransform}
H_{\mu {\rm w}}' =& \; \hbar \Omega_{\mu {\rm w}}{\rm cos}(\omega_{\mu {\rm w}} t -\phi_{\mu {\rm w}}) \big(e^{i(7 \omega_{\rm rf} + \alpha) t} \! \left | 4,4 \right \rangle \langle 3,3  | + e^{-i(7 \omega_{\rm rf} + \alpha) t}\left | 3,3 \right \rangle \langle 4,4  | \big). 
\end{align}\\
\noindent Multiplying through, dropping terms that oscillate at frequency $2\omega_{\mu{\rm w}}$, and substituting $\sigma_{x} = \hbar \big( \! \left | 4,4 \right \rangle \langle 3,3  | + \left | 3,3 \right \rangle \langle 4,4  | \big)$ and $\sigma_{y} = -i \hbar \big( \! \left | 4,4 \right \rangle \langle 3,3  | - \left | 3,3 \right \rangle \langle 4,4  | \big)$, we obtain the $\mu$w Hamiltonian in the rotating wave approximation, 

\begin{equation}
H_{\mu {\rm w}}' =\frac{\hbar\Omega_{\mu {\rm w}}}{2} \Big\lbrack {\rm cos}(\phi_{\mu {\rm w}})\sigma_{x} - {\rm sin}(\phi_{\mu {\rm w}})\sigma_{y} \Big\rbrack = H_{\mu{\rm w}}(\phi_{\mu{\rm w}}).
\end{equation}

\noindent Finally we pull it all together to obtain the control Hamiltonian,

\begin{align}
H_{\rm C} =&  \; H_0'+H_{\rm{rf}}^{(4)}(\phi_x,\phi_y)+H_{\rm{rf}}^{(3)}(\phi_x,\phi_y)+H_{\mu{\rm w}}(\phi_{\mu{\rm w}}) \nonumber \\
  =&  \; \Big\lbrack \frac{3 \hbar \Omega_{0}}{2}(1+g_{\rm rel}) - \frac{25 g_{\rm rel} \hbar^{2} \Omega_{0}^{2}}{2 \Delta E_{\rm HF}} - \frac{\hbar}{2}(\Delta_{\mu {\rm W}} - 7 \Delta_{\rm rf}) \Big\rbrack \big(P^{(4)}-P^{(3)}\big) \nonumber \\ 
&+ \hbar\Omega_{0}(1+g_{\rm rel})F_{z}^{(3)} + \frac{g_{\rm rel}\hbar^2 \Omega_{0}^{2}}{\Delta E_{\rm HF}}\big({F_{z}^{(4)}}^{2} - {F_{z}^{(3)}}^{2}\big) - \hbar\Delta_{\rm rf} \big(F_{z}^{(4)} - F_{z}^{(3)}\big)\nonumber \\
&+\frac{\hbar\Omega_{x}}{2} \Big\lbrack {\rm cos}(\phi_{x})F^{(4)}_{x} - {\rm sin}(\phi_{x})F^{(4)}_{y}\Big\rbrack + \frac{\hbar\Omega_{y}}{2}\Big\lbrack {\rm sin}(\phi_{y})F^{(4)}_{x}+{\rm cos}(\phi_{y})F^{(4)}_{y} \Big\rbrack \nonumber \\
& + \frac{g_{\rm rel}\hbar\Omega_{x}}{2} \Big\lbrack {\rm cos}(\phi_{x})F^{(3)}_{x} + {\rm sin}(\phi_{x})F^{(3)}_{y}\Big\rbrack + \frac{g_{\rm rel}\hbar\Omega_{y}}{2}\Big\lbrack -{\rm sin}(\phi_{y})F^{(3)}_{x}+{\rm cos}(\phi_{y})F^{(3)}_{y} \Big\rbrack \nonumber \\
& + \frac{\hbar\Omega_{\mu {\rm w}}}{2} \Big\lbrack {\rm cos}(\phi_{\mu {\rm w}})\sigma_{x} - {\rm sin}(\phi_{\mu {\rm w}})\sigma_{y} \Big\rbrack
\end{align}

\end{document}